\documentclass[a4paper,11pt]{article}
\usepackage{pos}
\usepackage{float}
\usepackage{caption}
\usepackage{booktabs}
\usepackage{multirow}

\newcommand{\be}{\begin{equation}}
\newcommand{\ee}{\end{equation}}

\title{Parton distributions in the SMEFT from high-energy
Drell-Yan tails}
\author*[a]{Maeve Madigan}
\author[a]{James Moore}

\affiliation[a]{DAMTP, University of Cambridge, Wilberforce Road, Cambridge, CB3 0WA, United Kingdom}

\abstract{
	We present a study of the interplay between PDF and EFT effects
	in high-mass Drell-Yan at the LHC.
	We quantify the impact of a consistent joint determination of the PDFs and Wilson coefficients on the bounds obtained on the EFT,
	and examine the effect on the PDFs, determining the extent to which EFT signals could be reabsorbed into the large-$x$ quark and anti-quark PDFs.
	Finally, we present dedicated projections for the High-Luminosity LHC and evaluate its ultimate potential to constrain the EFT parameters, while taking into account potential modifications of the proton structure.
	}

\FullConference{%
  *** The European Physical Society Conference on High Energy Physics (EPS-HEP2021), ***\\
  *** 26-30 July 2021 ***\\
  *** Online conference, jointly organized by Universität Hamburg and the research center DESY ***
}

\begin{document}
\maketitle

\section{Introduction}
The Standard Model (SM) has been tested to an unprecedented level by the Large Hadron Collider (LHC).
Searches for physics beyond the SM have placed strict constraints on new physics, 
indicating that new physics may lie at mass scales 
larger than
energies accessible to the LHC.
Indirect searches will provide 
a crucial avenue to new physics in this case,
searching for evidence in the form of
subtle distortions of cross sections and decay rates
relative to their SM predictions.
The Standard Model Effective Field Theory (SMEFT) 
provides a powerful theoretical framework for capturing
the indirect effects of new physics on LHC observables.

The high-mass tails of Drell-Yan (DY) invariant mass distributions 
are a particularly interesting indirect probe of new physics.  
The four-fermion operators of the dimension-6 SMEFT contribute at 
tree-level to the Drell-Yan amplitude, causing it to scale with 
energy as
$\mathcal{A} \propto E^2 / \Lambda^2$,  where $E$ is the energy of the process and $\Lambda$ is the new physics scale.
As a result, the SMEFT operators have a 
significant contribution to the high-mass bins of the DY distribution
and powerful constraints may be obtained. In fact,
high-mass DY measurements have been found to produce constraints 
on EFT operators which are competitive with constraints from lower energy measurements
of higher precision~\cite{Farina:2016rws}.  
Given this high energy scaling behaviour, 
nonzero values of the four-fermion operator coefficients
would result in a smooth distortion of the high-mass DY tail.
In order
to pin down this subtle sign of new physics,
it is important that we understand our
theoretical inputs and their associated uncertainties.

The PDFs form a key ingredient in the calculation of our
theoretical predictions for LHC observables.  
The PDFs are determined from fits to experimental data, including 
high-mass and high-$p_{T}$ measurements.  In particular, 
the large-$x$ region of quark and antiquark PDFs receives important constraints
from high-mass DY measurements.
However, PDFs are determined from fits to data under the assumption of 
the SM.
This leads to an inconsistency in our theoretical predictions for LHC observables: although
we may calculate a partonic cross section in the SMEFT, it is then convoluted
with a PDF which assumes the SM.
We will quantify the impact 
of this inconsistency on the bounds obtained on the EFT.  By performing a consistent simultaneous
determination of the EFT and PDF coefficients, we will quantify the extent to which the EFT bounds are modified.
We will also analyse the change in the PDFs to determine the extent to which it is possible to
reabsorb EFT effects into the PDFs.
A first take on this challenge was presented in a proof-of-concept study~\cite{Carrazza:2019sec} using
deep inelastic scattering (DIS) data.
Here
we extend this work to include LHC processes,
performing a joint determination of PDFs and EFT coefficients from both DIS and Drell-Yan data.
For a more detailed discussion we refer to our main work~\cite{Greljo:2021kvv}.

\section{Framework}
In this Section we briefly discuss the details of the data, theoretical predictions 
and fitting methodology used to produce a simultaneous determination of the EFT and PDFs.
We refer the reader to Ref.~\cite{Greljo:2021kvv} for a more detailed outline.

\paragraph{Data:}the present analysis is based on the DIS and DY measurements which were part of
the strangeness study of~\cite{Faura:2020oom}, which in turn was a variant of the
NNPDF3.1 global PDF determination~\cite{Ball:2017nwa}, extended with additional high-mass
DY cross sections.
There are 5 datasets in the high-mass category.
The highest invariant mass reach is provided
by the CMS 13 TeV measurement of the DY invariant mass 
distribution, with the 
highest bin reaching up to 3.0 TeV~\cite{Sirunyan:2018owv}.
No other datasets beyond DIS and DY are considered.

\paragraph{Theoretical predictions:}we formulate the SMEFT 
in terms of a simple but well-motivated BSM benchmark scenario: 
the $\hat{W}$ and $\hat{Y}$ electroweak oblique 
parameters generated in universal theories that modify the electroweak gauge boson propagators~\cite{Farina:2016rws}.
A second, flavour non-universal benchmark scenario is considered in Ref.~\cite{Greljo:2021kvv}.
At dimension-6 in the SMEFT, the effect of the $\hat{W}$ and $\hat{Y}$ parameters on DY observables
can be reproduced by a flavour universal linear combination of
four-fermion operators.
%In our main work we consider a second benchmark scenario~\cite{Greljo:2017vvb}, motivated by the evidence of lepton flavour universality (LFU) violation in $B$-meson decays recently reported by the LHCb collaboration~\cite{Aaij:2014ora, Aaij:2017vbb, Aaij:2019wad,Aaij:2021vac}.
%In Ref.~\cite{} we further consider a second, flavour-specific scenario
%motivated by the evidence of lepton flavour universality (LFU) violation in $B$-meson
%decays recently reported by the LHCb collaboration~\cite{Aaij:2014ora, Aaij:2017vbb, Aaij:2019wad,Aaij:2021vac}.
We calculate theoretical predictions for the SMEFT by augmenting the SM predictions with a K-factor,
\begin{equation}
	\mathrm{d} \sigma_{\rm SMEFT} = \mathrm{d} \sigma_{\rm{SM}} \times K_{\mathrm{SMEFT}}
\end{equation}
where  $\mathrm{d} \sigma_{\rm{SM}}$ denotes the SM predictions for the DIS and DY cross sections, calculated at NNLO in QCD and NLO in EW.  $K_{\mathrm{SMEFT}}$
is given by
\begin{equation}
	\begin{split}
		K_{\mathrm{SMEFT}} &= 1 + \hat{W} R_{\mathrm{SMEFT}}^{\hat{W}} + \hat{Y} R_{\mathrm{SMEFT}}^{\hat{Y}} \, ,\\ 
		R_{\mathrm{SMEFT}}^{\hat{W}} &\equiv \displaystyle \left( {\cal L}_{ ij}^{\rm NNLO} \otimes d\widehat{\sigma}_{ij,{\rm SMEFT}}^{\hat{W}}\right)
 \big/ \left( {\cal L}_{ ij}^{\rm NNLO} \otimes d\widehat{\sigma}_{ij,{\rm SM}} \right) \, ,
	\end{split}
\end{equation}
%${\cal L}_{ ij}^{\rm NNLO}$ denotes the partonic luminosity
%evaluated at NNLO QCD,
and $R_{\mathrm{SMEFT}}^{\hat{Y}}$ is defined similarly, where ${\cal L}_{ ij}^{\rm NNLO}$ denotes the partonic luminosity
evaluated at NNLO in QCD.  All SMEFT calculations are linear in $\hat{W}$, $\hat{Y}$, as shown.

\paragraph{Methodology:} 
we extract bounds on the EFT by performing a scan of the $(\hat{W}, \hat{Y})$ parameter space,  evaluating the $\chi^2$ test statistic at each point, where
\be
\chi^{2} = \sum_{i, j}^{n_{\mathrm{dat}}} (D_{i} - T_{i}) ( \mathrm{cov}^{-1})_{ij} (D_{j} - T_{j}) \, .
\ee
$T_{i}$ denote the theoretical predictions, $D_{i}$ are the central values of the experimental data and cov$_{ij}$ 
denotes the experimental covariance matrix.
Most EFT analyses make use of SM PDFs in this procedure, calculating $T_{i}(\hat{W}_{j}, \hat{Y}_{j})$ using the
same SM PDF set at each sampling point $(\hat{W}_{j}, \hat{Y}_{j})$.
To perform a consistent simultaneous determination,
at each sampling point $(\hat{W}_{j}, \hat{Y}_{j})$ we will instead calculate $T_{i}(\hat{W}_{j}, \hat{Y}_{j})$ using 
a SMEFT PDF which has been fit under the assumption of the SMEFT at that same point $(\hat{W}_{j}, \hat{Y}_{j})$.
Each PDF fit is performed using \texttt{NNPDF3.1}.  
Close enough to a local minimum,
the $\chi^2$ as a function of the EFT coefficients can be approximated by a quadratic form
from which we then extract bounds on the EFT parameters.

\section{Results from DIS and high-mass DY data}
%%%%%%%%%%%%%%%%%%%%%%%%%%%%%%
%%%%%%%%%%%%%%%%%%%%%%%%%%%%%%
\begin{figure}[h]
\begin{center}
  \includegraphics[width=0.49\textwidth]{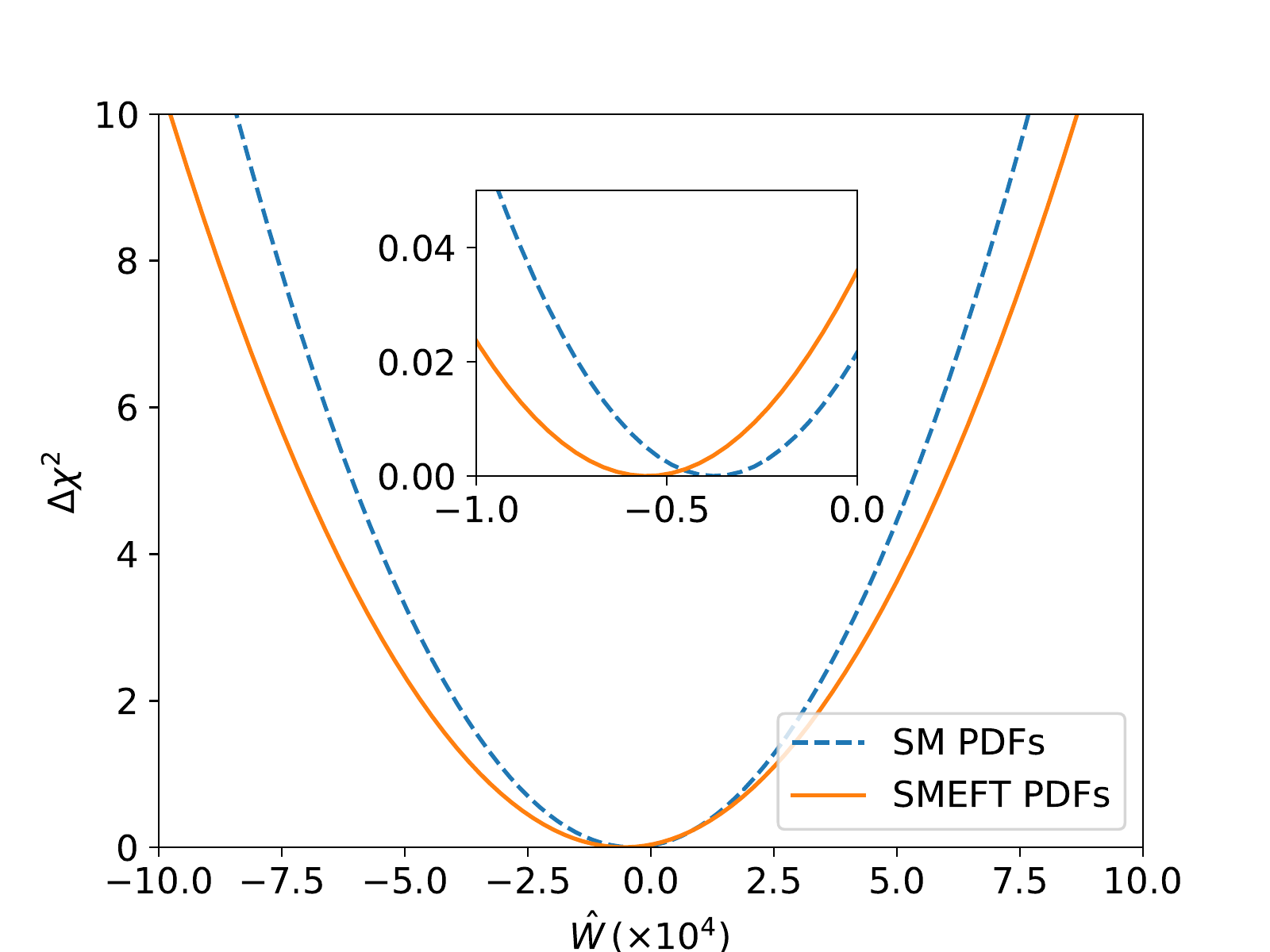}
    \includegraphics[width=0.49\textwidth]{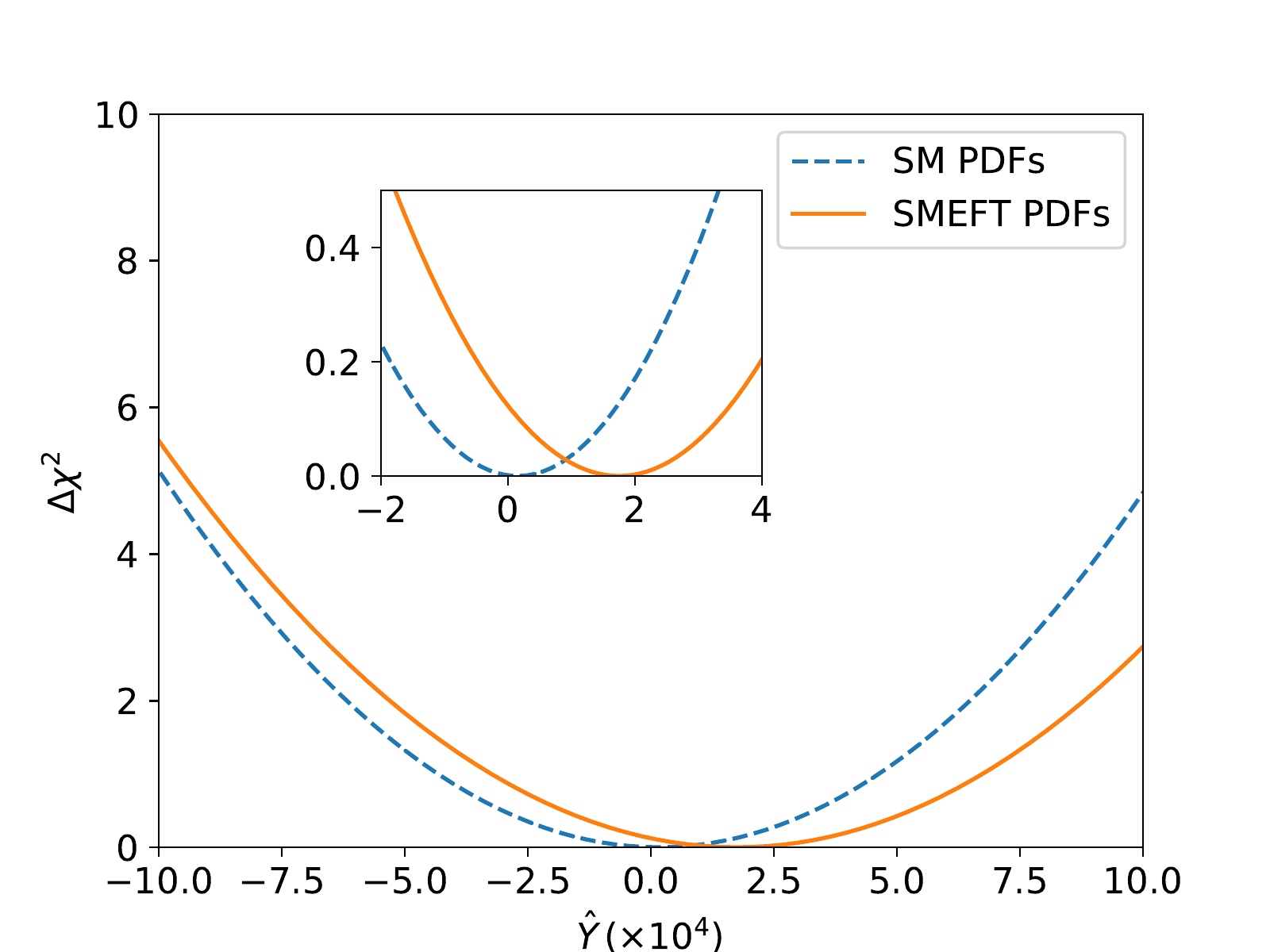}
    \caption{\label{fig:parabolas2} Comparison
      between the results of the parabolic fits to $\Delta\chi^2$
      for the $\hat{W}$ (left) and $\hat{Y}$ (right panel) parameters for either the SMEFT PDFs
	or
      the SM PDFs.
      The insets zoom in on the region close to $\Delta\chi^2\simeq 0$.
    }
\end{center}
\end{figure}
%%%%%%%%%%%%%%%%%%%%%%%%%%%%%%
%%%%%%%%%%%%%%%%%%%%%%%%%%%%%%
We begin by discussing the impact of the simultaneous fit 
on the bounds on the $\hat{W}$ and $\hat{Y}$ parameters, taking one parameter at a time
while setting the other to zero.
Fig.~\ref{fig:parabolas2} compares the results
of the parabolic fits to $\Delta \chi^2$ based on the SMEFT PDFs
with their
counterparts
obtained in the case of the SM PDFs.
The insets highlight the region close to $\Delta\chi^2\simeq 0$.
For the $\hat{W}$ parameter, the consistent use of SMEFT PDFs leaves
the best-fit value essentially unchanged but increases the coefficient
uncertainty $\delta \hat{W}$, leading to a
broader parabola.
Similar observations can be derived for the $\hat{Y}$ parameter, though here
one also finds an upwards shift in the best-fit values by $\Delta\hat{Y}\simeq 2\times 10^{-4}$
in addition to a parabola broadening, when SMEFT PDFs are
consistently used.
We note that the SM PDF parabolas in Fig.~\ref{fig:parabolas2} are evaluated
using the central PDF replica and hence do not account for PDF uncertainties.
Table~\ref{tab:bound1w} summarises
the corresponding 95\% CL bounds on the $\hat{W}$ and $\hat{Y}$
parameters obtained 
using either the SM or the SMEFT PDFs shown in Fig.~\ref{fig:parabolas2}.
Here, we produce the bounds obtained from SM PDFs without (upper) and
with (lower entry) accounting for PDF uncertainties.
%

%%%%%%%%%%%%%%%%%%%%%%%%%%%%%%%%%%%%%%%%%%%%%%%%%%%%%%%%%%%%%%%%%%%%%%%%%%%%%%%%%%
\begin{table}[h]
  \renewcommand{\arraystretch}{1.40}
  \centering
  \begin{tabular}{|c|c|c|c|c|c|}
  \toprule
    Parameter & SM PDFs  & SMEFT PDFs  & Parameter  & SM PDFs  & SMEFT PDFs \\
    \toprule
	  \multirow{2}{*}{$\hat{W}\times 10^4$}  & $[-5.5, 4.7]$ & \multirow{2}{*}{ $[-6.4, 5.3] $} & \multirow{2}{*}{$\hat{Y}\times 10^4$} & $[-8.8, 9.2] $  & \multirow{2}{*}{$[-8.3, 11.8]$} \\
	   &  $[-6.8, 6.3]$ &  &  & $[-11.1, 12.0] $ &  \\ 
    \bottomrule
  \end{tabular}
  \caption{\label{tab:bound1w} \small The 95\% CL bounds on the $\hat{W}$ and $\hat{Y}$
    parameters obtained from the  corresponding parabolic fits to
    the $\Delta\chi^2$ values calculated from  either the SM or the SMEFT PDFs.
    For the SM PDF results, we indicate the bounds obtained without (upper)
    and with (lower entry) PDF uncertainties accounted for; the SMEFT PDF
    bounds already include  PDF uncertainties by construction.
}
\end{table}
%%%%%%%%%%%%%%%%%%%%%%%%%%%%%%%%%%%%%%%%%%%%%%%%%%%%%%%%%%%%%%%%%%%%%%%%%%%]

\begin{figure}[h]
\begin{center}
\includegraphics[width=0.47\textwidth]{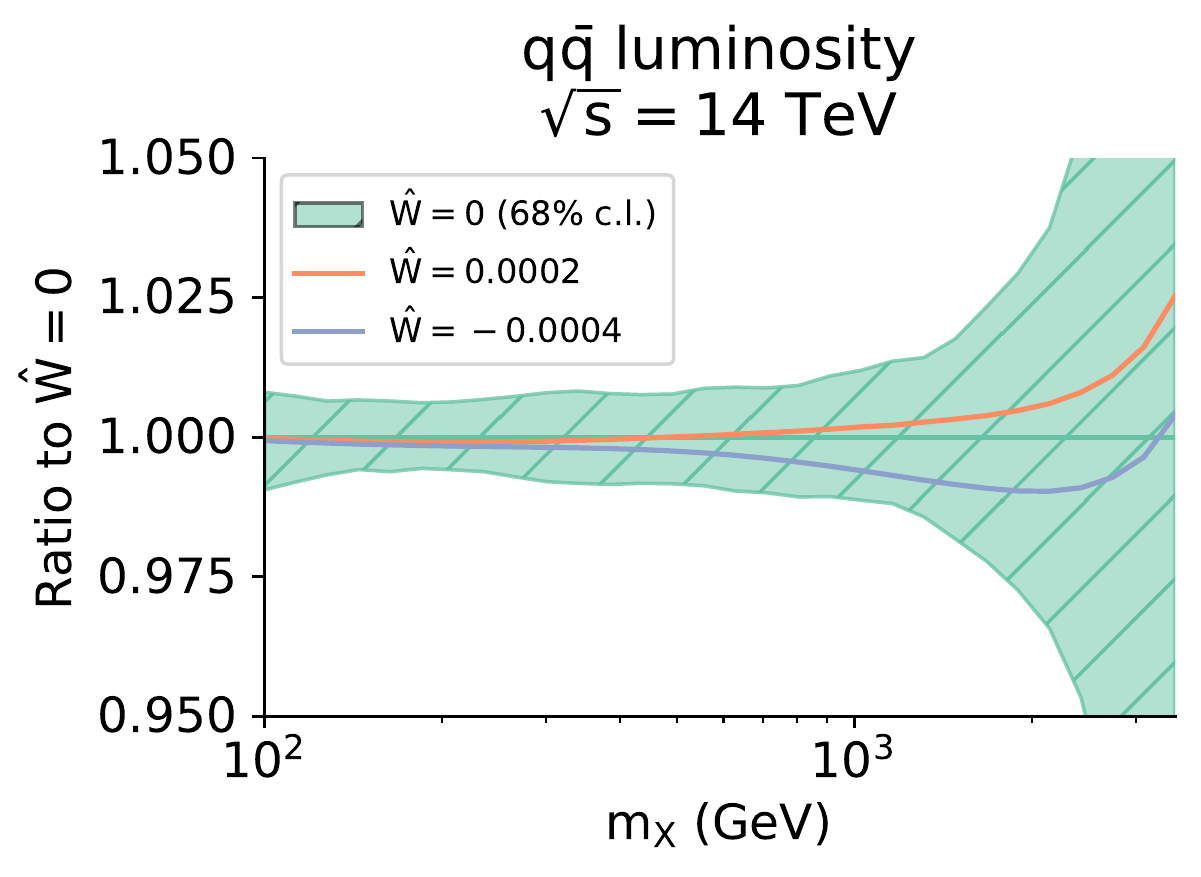}
\includegraphics[width=0.47\textwidth]{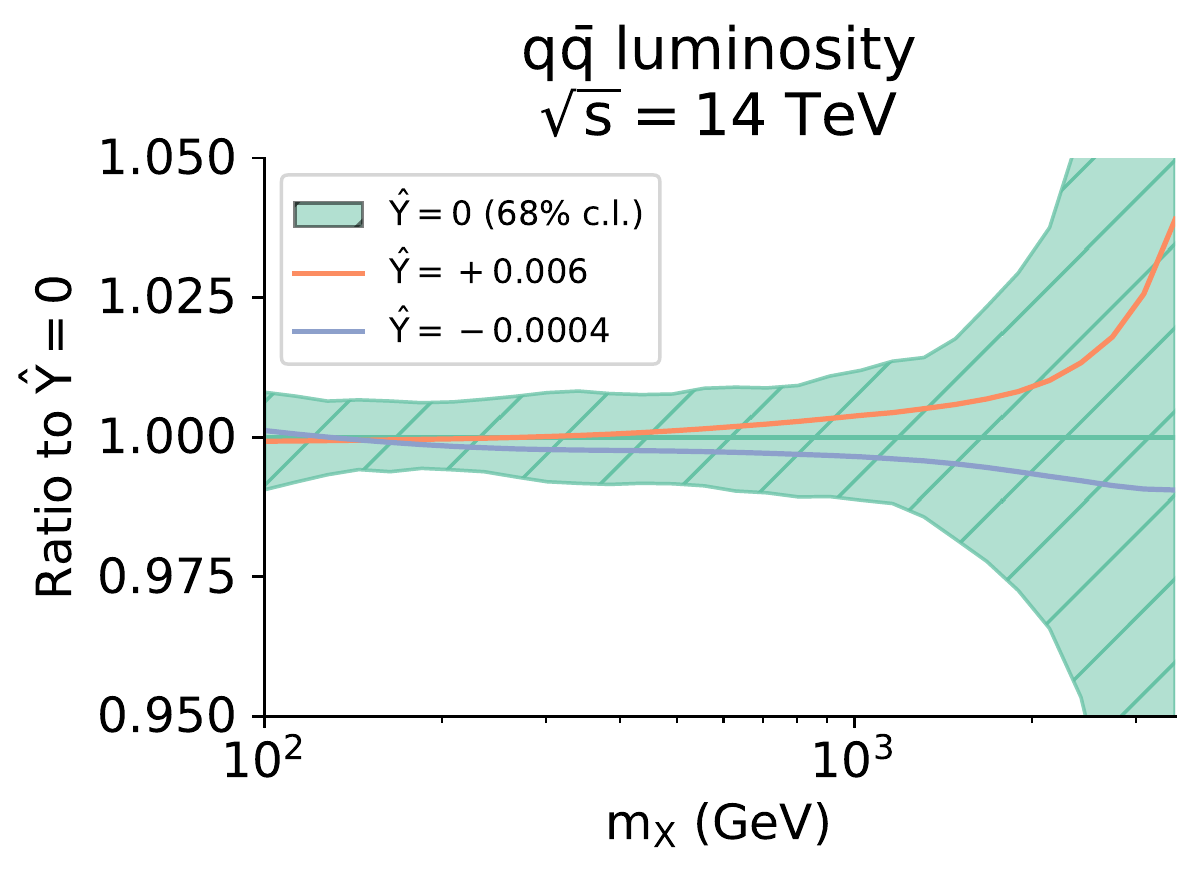}
\caption{\label{fig:SMEFT_lumis}
	Comparison between the SM PDF quark-antiquark luminosity with its SMEFT counterparts,
	displayed as ratios to the central value of the SM quark-antiquark luminosity,
	for representative values of the $\hat{W}$ (left) and $\hat{Y}$ (right)
	parameters.}
\end{center}
\end{figure}

Next, we turn to the effect of the simultaneous EFT and PDF determination on the
PDFs.
In Fig~\ref{fig:SMEFT_lumis} we compare the SM and SMEFT quark-antiquark luminosities
at representative values of $(\hat{W}, \hat{Y})$, choosing the values to be within
 the upper and lower limits of the 95\% CL intervals reported in Table~\ref{tab:bound1w}.
 The luminosities are displayed as ratios to the central values of the SM $q \bar{q}$ luminosity,
 for which we also display the luminosity uncertainty (green).
 The plots indicate that
the EFT-induced shifts on the luminosity are smaller than
its standard deviation. 
Together with the small shift in bounds shown in Fig~\ref{fig:parabolas2} and Table~\ref{tab:bound1w}, 
this shows that
with current data, the interplay between EFT effects and PDFs in the high-mass Drell-Yan tails is appreciable but remains subdominant as compared to other sources of uncertainty.
%%%%%%%%%%%%%%%%%%%%%%%%%%%%%%%%%%%%%%%%%%%%%%%%%%%%%%%%%%%%%%%%%%%%%%%%
%%%%%%%%%%%%%%%%%%%%%%%%%%%%%%%%%%%%%%%%%%%%%%%%%%%%%%%%%%%%%%%%%%%%%%%%

\section{High-luminosity projections}
Having found a moderate interplay between the EFT and PDFs using data
from DIS and DY processes, we can expect that this interplay may become
significant as more data becomes available.
With this motivation, we repeat the joint PDF and EFT determination 
of the previous section, this time including projected pseudodata from the
High-Luminosity LHC (HL-LHC).  Following the strategy of Ref.~\cite{Khalek:2018mdn}
we generate pseudodata for neutral current (NC) and charged current (CC) DY. 
%
%The pseudodata are binned in invariant mass $m_{\ell \ell}$ and transverse mass $m_{T}$ respectively,
%and we restrict to bins satisfying $m_{\ell \ell}$, $m_{T} > 500$ GeV.
%We adopt reference measurements from ATLAS~\cite{Aad:2019wvl} and CMS~\cite{Sirunyan:2018owv},
%modelling aceptance cuts and systematic uncertainties from these.
%Systematic uncertainties are reduced by a factor of $f_{\mathrm{red}} = 0.2$ to
%account for the expected reduction in systematic errors foreseen at the HL-LHC. 
%
The inclusion of CC DY data lifts a flat direction in $(\hat{W}, \hat{Y})$-space,
allowing a simultaneous determination of both the $(\hat{W}, \hat{Y})$
parameters
and the PDFs.

\begin{figure}[h!]
\begin{center}
  \includegraphics[width=0.49\textwidth]{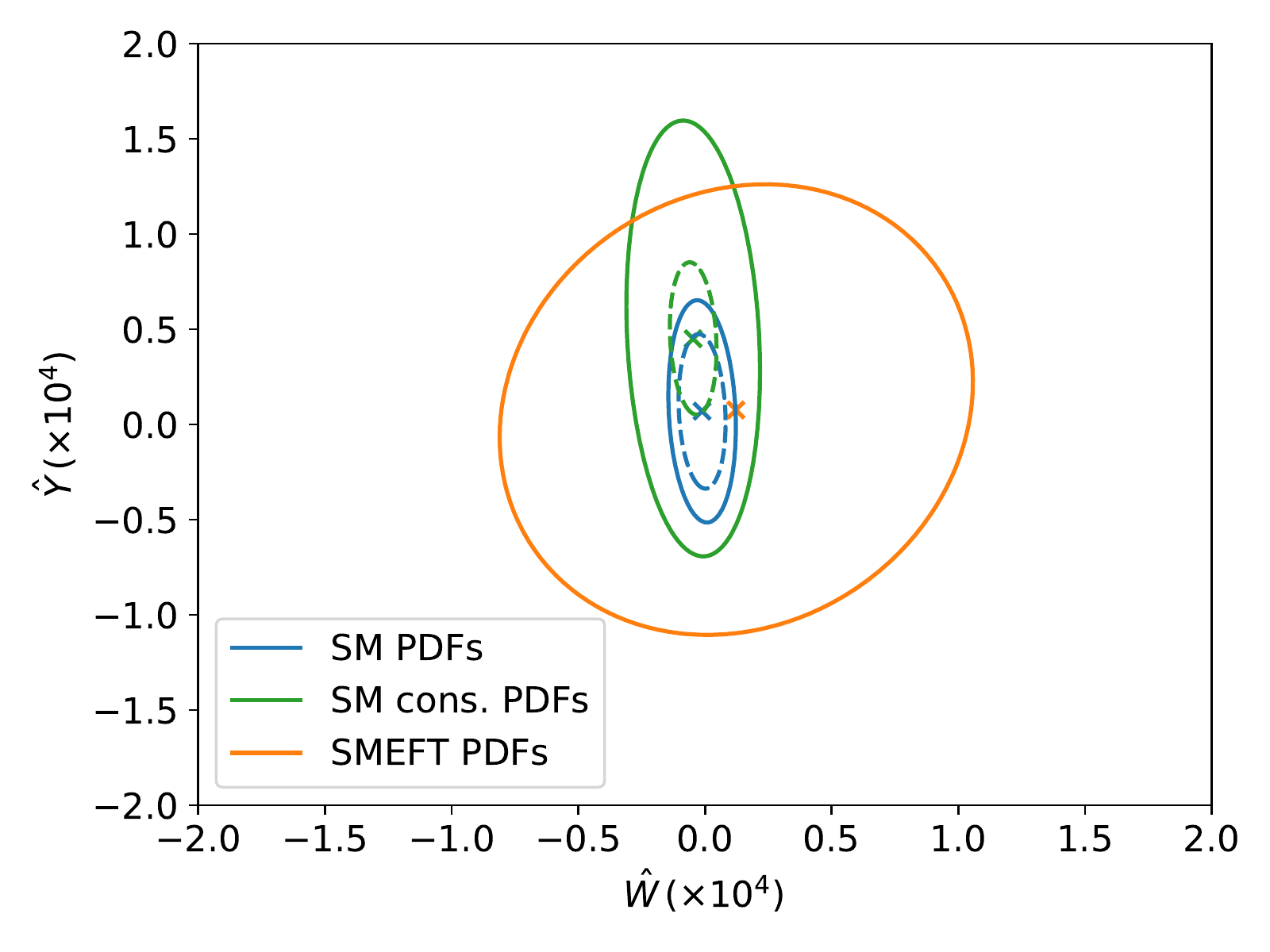}
  \includegraphics[width=0.49\textwidth]{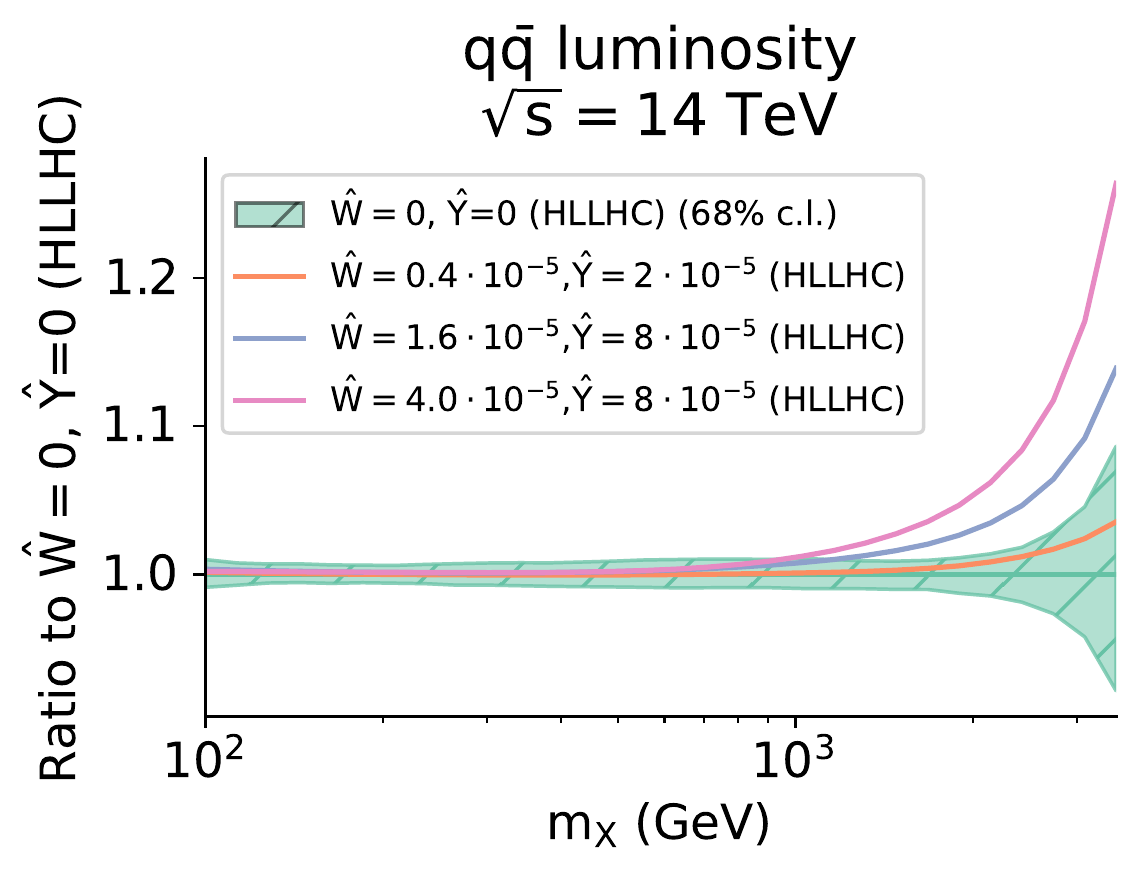}
  \caption{\label{fig:hllhc-ellipse} On the left, we display the 95\% confidence level contours
    in the ($\hat{W}$,$\hat{Y}$) plane obtained from the DIS+DY fits that
    include the high-mass Drell-Yan HL-LHC pseudo-data 
    when using either SM PDFs (blue) or conservative SM PDFs (green). 
	PDF
    uncertainties are included in the solid lines and not included in
    the dashed lines.  The results are compared to
    those obtained in a simultaneous fit, namely with SMEFT PDFs
    (orange). 
        The right plot compares the quark-antiquark SM PDF luminosity in the fits including the HL-LHC pseudo-data
    with those obtained in the SMEFT PDF fits for representative values
    of the $\hat{W}$ and $\hat{Y}$ parameters.
        }
\end{center}
\end{figure}

We find that including the HL-LHC pseudodata in a fit of PDFs and
in a fit of SMEFT coefficients while neglecting their interplay
could result in a significant underestimate
of the uncertainties associated to the EFT parameters, as shown in Table~\ref{tab:hlbounds}.
The marginalised 95\% CL bound on each of the $\hat{W}$ and $\hat{Y}$ parameters becomes looser
once SMEFT PDFs are consistently used,
even once PDF uncertainties are fully accounted for.
These results are graphically displayed in
Fig.~\ref{fig:hllhc-ellipse} (left), where the 95\% confidence level contours
    in the ($\hat{W}$,$\hat{Y}$) plane obtained 
    when using either SM PDFs (blue)
    or SMEFT PDFs (orange) are compared. All solid countours include PDF
    uncertainties.
    The dashed contours that do not include PDF
    uncertainties are also indicated to visualise the impact of the
    inclusion of the PDF uncertainties.

%%%%%%%%%%%%%%%%%%%%%%%%%%%%%%%%%%%%%%%%%%%%%%%%%%%%%%%
\begin{table}[h!]
 \renewcommand{\arraystretch}{1.40}
  \centering
  \begin{tabular}{l|c|c|c}
          & SM PDFs & SM conservative PDFs & SMEFT PDFs  \\
    \hline
          %\multirow{2}{*}{ $\hat{W}\times 10^5$ (68\% CL)} & $[-0.7, 0.5]$ & $[-1.0, 0.0]$ & \multirow{2}{*}{$[-4.5, 6.9]$}  \\
      %& $[-1.0, 0.9]$ & $[-4.0, 2.8]$ &  \\
    %\midrule
    \multirow{2}{*}{$\hat{W}\times 10^5$} & $[-1.0, 0.8]$ & $[-1.4, 0.4]$ & \multirow{2}{*}{$[-8.1, 10.6]$}\\
      & $[-1.4, 1.2]$ &  $[-4.3, 3.1]$ &   \\
   \hline
   %\multirow{2}{*}{$\hat{Y}\times 10^5$ (68\% CL)} & $[-1.8, 3.2]$ & $[2.1, 7.0]$ & \multirow{2}{*}{$[-6.4, 8.0]$}  \\
      % & $[-3.7, 4.7] $ & $[-3.4,11.2]$ &  \\
   %\midrule
   \multirow{2}{*}{$\hat{Y}\times 10^5$} & $[-3.4, 4.7] $ & $[0.5, 8.5]$ & \multirow{2}{*}{$[-11.1, 12.6]$} \\
   & $[-5.3, 6.3] $ & $[-5.0, 13.7]$ &  \\
    \hline
  \end{tabular}
  \caption{ \label{tab:hlbounds}\small
    The 95\% CL
    marginalised bounds on the $\hat{W}$ and $\hat{Y}$
    parameters obtained from the two-dimensional ($\hat{W}$,$\hat{Y}$)
    fits that include the HL-LHC pseudo-data for NC and CC
    Drell-Yan distributions.
    As in Table~\ref{tab:bound1w},
    for the SM PDFs we indicate the bounds obtained without (upper)
    and with (lower entry) PDF uncertainties accounted for.
}
\end{table}
%%%%%%%%%%%%%%%%%%%%%%%%%%%%%%%%%%%%%%%%%%%%%%%%%%%%%%%

%%%%%%%%%%%%%%%%%%%%%%%%%%%%%%%%%%%%%%%%%%%%%%%%
%%%%%%%%%%%%%%%%%%%%%%%%%%%%%%%%%%%%%%%%%%%%5

The impact of the simultaneous determination on
%by examining the effects on 
the quark-antiquark luminosity is shown in Fig~\ref{fig:hllhc-ellipse} (right),
comparing the luminosities from SM PDFs to those
obtained from SMEFT PDFs for representative values of
$(\hat{W}, \hat{Y})$.
We find that the central value of the SMEFT PDF luminosity shifts greatly relative
to the SM PDFs, well outside the
one-sigma error band of the SM PDFs, while the
PDF uncertainties themselves are unchanged.
This change in central value of the large-$x$ PDFs partially reabsorbs the effects in the
partonic cross section induced by the SMEFT operators and leads to better
$\chi^2$ values as compared to those obtained with the SM PDFs.

\subsection*{Conservative PDF sets}

A further important question is whether the bounds obtained with SM
PDFs appearing on the leftmost column of Table~\ref{tab:hlbounds}  would become
more comparable to those obtained from the simultaneous fit of PDFs
and SMEFT coefficients, in case a
conservative set of PDFs was used in the analysis based on SM PDFs. 
A conservative PDF set is one which is obtained from data under the assumption of
the SM, but does not include any of the high-mass Drell-Yan sets (neither the
HL-LHC projections or the existing high-mass measurements from the LHC).
To
address this question, in Table~\ref{tab:hlbounds} we also display the
bounds that are obtained using a conservative PDF set.
We observe that, once this set of conservative PDFs is used as an input
PDF set and the PDF uncertainty is included in the computation of the EFT
bounds, the bounds widen as compared to the bounds computed using SM PDFs,
as shown in the central column of Table~\ref{tab:hlbounds}.
As a result, the size of the bounds obtained
by using conservative SM PDFs is closer to the size obtained from the
simultaneous fits, although still slightly underestimated.

Before we conclude, we note that the pseudodata generated for the purpose of
the HL-LHC projections
are generated under the assumption of the SM.
It is worth considering a more optimistic scenario in which 
new physics is present in the form of nonzero values of the $\hat{W}$ and $\hat{Y}$ parameters.
We generate HL-LHC pseudodata assuming $(\hat{W}, \hat{Y}) = (4, 8) \times 10^{-5}$, taking these 
values from within the 95 \% CL bounds found in Section 3, and recalculate the EFT bounds.
%%%%%%%%%%%%%%%%%%%%%%%%%%%%%%%%%%%%%%%%%%%%%%%%%%%%%%%
\begin{table}[h!]
 \renewcommand{\arraystretch}{1.40}
  \centering
  \begin{tabular}{l|c|c|c}
          & SM PDFs & SM conservative PDFs & SMEFT PDFs  \\
    \hline
	  $\hat{W}\times 10^5$ & $[-1.5, 1.2]$ & $[3.1, 5.0]$ & $[-5.3, 9.0]$\\
   \hline
	  $\hat{Y}\times 10^5$ & $[-3.1, 8.7] $ & $[5.8, 13.6]$ & $[-0.2, 26.7]$\\
    \hline
  \end{tabular}
  \caption{ \label{tab:hlboundsNP}\small
    We inject a spurious signal of new physics into the HL-LHC pseudodata, taking $(\hat{W}, \hat{Y}) = (4, 8) \times 10^{-5}$ as a benchmark. The table shows the 95\% CL
    marginalised bounds on the $\hat{W}$ and $\hat{Y}$
    parameters obtained from the two-dimensional ($\hat{W}$,$\hat{Y}$)
    fits that include this HL-LHC pseudodata.  PDF uncertainties are accounted for.
}
\end{table}
%%%%%%%%%%%%%%%%%%%%%%%%%%%%%%%%%%%%%%%%%%%%%%%%%%%%%%%
The results are shown in Table~\ref{tab:hlboundsNP}.  We find that the fully simultaneous fit 
does a good job of detecting new physics, with the bounds moving to the right relative to those in Table~\ref{tab:hlbounds}.
In contrast, the fit using SM PDFs that have seen the SMEFT-affected data are unable to detect new physics: the point $(\hat{W}, \hat{Y}) = (4, 8) \times 10^{-5}$ 
lies outside of the marginalised bounds at 95 \% CL shown in the leftmost column of Table~\ref{tab:hlboundsNP}.
Finally we find that using conservative SM PDFs we are able to detect the new physics, and the bounds are in fact tighter than 
those obtained using SMEFT PDFs.  Our results suggest that a more careful study of conservative PDFs will be very important in the future, as PDF fits continue to include more and more data, some of which could be SMEFT-contaminated. In particular, it will be crucial for those performing SMEFT fits to know whether a fully simultaneous PDF-SMEFT fit is required, or whether they can reliably use conservative sets instead.

\section{Conclusions}
Exploiting the full potential of current and future precision measurements at the LHC for indirect BSM searches requires the development of novel data interpretation frameworks capable of handling the interplay between PDF and EFT effects in the high energy tails of LHC distributions.  We present a study of this interplay using high-mass DY data, building on a previous study of DIS data~\cite{Carrazza:2019sec}.  We find that at present, the interplay between PDF and EFT effects remains moderate and subdominant relative to other sources of PDF uncertainty.  This situation will change, however, as we move towards the HL-LHC: our projections show that neglecting this interplay may lead to artificially precise bounds on the EFT.
A more detailed investigation into the definition of conservative PDFs is needed, particularly as PDF and EFT fits include more data, some of which may contain signs of new physics.  In parallel, it is crucial that we begin to develop a more powerful methodology capable of handling further SMEFT operators in order to move towards a truly global simultaneous fit.

\bibliographystyle{JHEP}
\bibliography{paperDYproc}

\providecommand{\href}[2]{#2}\begingroup\raggedright\begin{thebibliography}{1}

\bibitem{Farina:2016rws}
M.~Farina, G.~Panico, D.~Pappadopulo, J.~T. Ruderman, R.~Torre, and A.~Wulzer,
  {\it {Energy helps accuracy: electroweak precision tests at hadron
  colliders}},  {\em Phys. Lett.} {\bf B772} (2017) 210--215,
  [\href{http://arxiv.org/abs/1609.08157}{{\tt arXiv:1609.08157}}].

\bibitem{Greljo:2021kvv}
A.~Greljo, S.~Iranipour, Z.~Kassabov, M.~Madigan, J.~Moore, J.~Rojo, M.~Ubiali,
  and C.~Voisey, {\it {Parton distributions in the SMEFT from high-energy
  Drell-Yan tails}},  {\em JHEP} {\bf 07} (2021) 122,
  [\href{http://arxiv.org/abs/2104.02723}{{\tt arXiv:2104.02723}}].

\bibitem{Carrazza:2019sec}
S.~Carrazza, C.~Degrande, S.~Iranipour, J.~Rojo, and M.~Ubiali, {\it {Can New
  Physics hide inside the proton?}},  {\em Phys. Rev. Lett.} {\bf 123} (2019),
  no.~13 132001, [\href{http://arxiv.org/abs/1905.05215}{{\tt
  arXiv:1905.05215}}].

\bibitem{Faura:2020oom}
F.~Faura, S.~Iranipour, E.~R. Nocera, J.~Rojo, and M.~Ubiali, {\it {The
  Strangest Proton?}},  {\em Eur. Phys. J. C} {\bf 80} (2020), no.~12 1168,
  [\href{http://arxiv.org/abs/2009.00014}{{\tt arXiv:2009.00014}}].

\bibitem{Ball:2017nwa}
{\bf NNPDF} Collaboration, R.~D. Ball et~al., {\it {Parton distributions from
  high-precision collider data}},  {\em Eur. Phys. J.} {\bf C77} (2017), no.~10
  663, [\href{http://arxiv.org/abs/1706.00428}{{\tt arXiv:1706.00428}}].

\bibitem{Sirunyan:2018owv}
{\bf CMS} Collaboration, A.~M. Sirunyan et~al., {\it {Measurement of the
  differential Drell-Yan cross section in proton-proton collisions at $
  \sqrt{\mathrm{s}} $ = 13 TeV}},  {\em JHEP} {\bf 12} (2019) 059,
  [\href{http://arxiv.org/abs/1812.10529}{{\tt arXiv:1812.10529}}].

\bibitem{Khalek:2018mdn}
R.~Abdul~Khalek, S.~Bailey, J.~Gao, L.~Harland-Lang, and J.~Rojo, {\it {Towards
  Ultimate Parton Distributions at the High-Luminosity LHC}},  {\em Eur. Phys.
  J. C} {\bf 78} (2018), no.~11 962,
  [\href{http://arxiv.org/abs/1810.03639}{{\tt arXiv:1810.03639}}].

\end{thebibliography}\endgroup

\end{document}